\documentclass[12pt]{article}
\usepackage{xparse}
\usepackage{etoolbox}
\usepackage{color,xcolor}
\usepackage{amssymb,amsmath,bm,bbm}
\usepackage{epsf}
\usepackage{epsfig}
\usepackage{afterpage}
\usepackage{longtable}
\usepackage{cite}
\usepackage{latexsym, mathrsfs}
\usepackage{graphics}
\usepackage{url}
\usepackage{paralist}
\usepackage{bbold}
\usepackage{cleveref}
\usepackage{slashed}
\usepackage{adjustbox}
\usepackage{makecell}

\usepackage{braket}

\newcommand{\qq}{\quad \quad}

\newcommand{\be}{\begin{equation}}
\newcommand{\ee}{\end{equation}}
\newcommand{\bea}{\begin{eqnarray}}
\newcommand{\eea}{\end{eqnarray}}
\newcommand{\nn}{\nonumber}
\newcommand{\dd}{\displaystyle}

\newcommand{\gev}{\textrm{GeV}}
\newcommand{\mev}{\textrm{MeV}}

\newcommand{\noi}{\noindent}

\begin{document}

\begin{flushright} {BARI-TH/23-745}\end{flushright}

\medskip

\begin{center}
{\Large Nonleptonic $B_c$ decay rates from model independent relations}
\\[1.0 cm]
{Nicola~Losacco$^{a,b}$
\\[0.5 cm]}
{\small $^a$Dipartimento Interateneo di Fisica ``M. Merlin'', Universit\`a  e Politecnico di Bari, \\ via Orabona 4, 70126 Bari, Italy \\[0.1 cm] $^b$Istituto Nazionale di Fisica Nucleare, Sezione di Bari,  Via Orabona 4, I-70126 Bari, Italy
}
\end{center}

\vskip 0.8cm

\begin{abstract}
\noindent
Nonleptonic $B_c$ decays to $P$-wave charmonia and light  $\pi^+$, $K^+$, and $\rho^+$,  $K^{*+}$ mesons are analysed using factorization. The  hadronic form factors parametrizing the $B_c \to \chi_{cJ}(h_c)$ matrix elements are expressed in terms of universal functions at the leading order of an  expansion in the relative velocity of the heavy quarks in the $B_c$ rest-frame and in $1/m_Q$. Several ratios of branching fractions are evaluated, and when experimental information can be used, single branching fractions are presented. Both the $1P$ and $2P$ charmonia are considered. If the exotic candidate state $\chi_{c1}(3872)$ corresponds to $\chi_{c1}(2P)$, it should be produced in nonleptonic $B_c$ decays with predicted abundances with respect to the other states in the charmonium $2P$ spin four-plet.
 
\end{abstract}

\thispagestyle{empty}

\section{Introduction}

The decays of the $B_c$ meson, a hadron with the quarkonium structure and only weak transitions, represent a unique opportunity to analyse the properties of strong and weak interactions. The semileptonic decays are induced  by both beauty and charm quark transitions, and give access to the $V_{cb}$, $V_{ub}$, and $V_{cs}$,  $V_{cd}$ elements of the Cabibbo-Kobayashi-Maskawa (CKM) mixing matrix. Several studies have been devoted to them,  namely to  the processes with excited states \cite{Ebert:2010zu,Rui:2017pre,Wang:2018duy,Rui:2018kqr,Colangelo:2022awx}.
 The  $b \to c$ transitions, considered in the present analysis, involve $\lvert V_{cb} \rvert$ for which a  tension between determinations from inclusive and exclusive  $B$ decays is under scrutiny \cite{Gambino:2020jvv,Colangelo:2016ymy}. Properties of the Standard Model (SM) as lepton flavour universality can be tested in semileptonic $B_c$ decays \cite{Tran:2018kuv,Cohen:2018dgz,Leljak:2019eyw,Tang:2022nqm,Colangelo:2021dnv}.
 Rare semileptonic $B_c$  modes induced by  flavour changing neutral current transitions of both beauty and charm quark also probe the SM and its extensions
 \cite{Colangelo:2021myn}. In such processes the hadronic uncertainty is in the $B_c$ form factors parametrizing the matrix elements of current operators.

For nonleptonic $B_c$ decays the theoretical description is more difficult than for the semileptonic modes, since it deals with the hadronic matrix elements of four-quark operators.
The factorization approach represents a simple way to treat such operators in the decay amplitudes. For $b \to c \bar q_1 q_2$ and neglecting the annihilation topology, the matrix elements of the four-quark operators is factorized in the product of two terms, one describing the $W$ production of the light meson and parametrized by the decay constant, the other one the $B_c$ to charmonium current  matrix element.\footnote{ Examples of such calculations are in \cite{Lusignoli:1990ky,Ivanov:2006ni}.} Invoking the heavy quark spin symmetry, it has been observed that relations can be established among the $B_c$ to charmonium form factors,  namely  the two $B_c\to S$-wave  $(\eta_c,J/\psi)$ modes, the four  $B_c\to P$-wave charmonium $(h_c,\chi_{cJ})$ modes, etc., in a  region of the phase space close to  zero recoil \cite{Jenkins:1992nb,Colangelo:1999zn}. Such relations have been worked out through an expansion in the relative three-velocity of the heavy quarks in $B_c$ and in $1/m_Q$, including NLO terms
\cite{Colangelo:2022lpy,Colangelo:2022awx}. They can constrain the hadronic uncertainties and  establish  model-independent connections among different processes.
 
The present study is focused on nonleptonic  $B_c$  decays to $1P$ and $2P$ charmonium plus a light meson $M$: $B_c \to \chi_{cJ}(h_c) M$. They are of interest for several reasons.  For example,  the analysis of new physics contributions in processes like  $B_c^+ \to K^+ \, K^- \, \pi^+$ must consider  $B_c^+ \to \chi_{c0} \, \pi^+$ as an intermediate state \cite{Chen:2021vmb}. Moreover, the $B_c$ nonleptonic transitions   probe the structure of states as $\chi_{c1}(3872)$ (previously denoted  as $X(3872)$), whose multiquark/conventional quarkonium nature 
is under investigation \cite{Brambilla:2019esw,Maiani:2022psl,COLANGELO2007166,Achasov:2019cfe,Achasov:2022puf,Chen:2016qjuRefRed,Guo:2017jvcRefRed,Meng:2022ozqRefRed}. The nonleptonic $B_c$ processes  provide additional information  with respect to the semileptonic ones \cite{Wang:2007sxa,WANG2016261}. Of  particular interest is to exploit the model independent relations among the  $B_c \to P$-wave charmonium form factors  \cite{Colangelo:2022awx}.

The plan of the paper is as follows. In Sec.~\ref{nonleptonicTh}, using  the effective Hamiltonian inducing the $b \to c \bar q_1 q_2$ transition with $q_{1,2}$ light quarks, 
 we evaluate the decay widths of  $B_c$ to $P$-wave charmonium and a light pseudoscalar/vector meson. In Sec.~\ref{Observables} we use the LO  relations for the form factors in \cite{Colangelo:2022awx} to predict several ratios of branching fractions. In Sec.~\ref{LHCb} we comment on  a recent measurement of the ratio  
$ \dd \frac{\mathcal{B}(B_c^+ \to B_s^0 \pi^+)}{\mathcal{B}(B_c^+ \to J/\psi \pi^+)}$ 
 \cite{LHCb:2022htj}, suggesting a violation of factorization in  the $B_c^+ \to B_s^0 \pi^+$ mode. Then we conclude.

\noindent 

\section{Nonleptonic $B_c$ decays in the factorization approach}\label{nonleptonicTh}

The effective Hamiltonian governing the nonleptonic  $B_c$ decays to  charmonium and a light meson is given by \cite{Neubert:1997uc,Neubert:2000ag}

\be
\mathcal{H}_{eff}= \frac{G_F}{\sqrt{2}}V_{c b}^* V_{u q}\left(C_1 (\mu) Q_1 (\mu) + C_2 (\mu) Q_2 (\mu) \right) + {\text h.c.} \label{Hamil}
\ee
\noindent where 
\bea
Q_1 &=& \bar{u}_{\alpha} \gamma^{\mu} (1-\gamma_5) q_{\alpha} \bar{b}_{\beta} \gamma_{\mu} (1-\gamma_5) c_{\beta} \nn \\
Q_2 &=& \bar{u}_{\alpha} \gamma^{\mu} (1-\gamma_5) q_{\beta} \bar{b}_{\beta} \gamma_{\mu} (1-\gamma_5) c_{\alpha} \, .
\eea
$\alpha, \beta$ are color indices, and 
$q$ identifies the down type quark of the final light meson.  $\mu$ represents the scale dependence of the Wilson coefficients which encode the short-distance physics from energy greater than $\mu$. It is cancelled in the amplitude by the $\mu$ dependence of the operators matrix elements. In our computation of $B$ decays we are setting $\mu = m_b$. Strong interaction effects are taken into account considering the Wilson coefficient $C_2(\mu) \neq 0$.
Using the Fierz color rearrangement  relation 
\be
\delta_{\alpha \lambda} \delta_{\rho \beta} = \frac{1}{C} (T_a)_{\alpha \beta}(T_a)_{\rho \lambda} + \frac{1}{N_c} \delta_{\alpha \beta} \delta_{\rho \lambda},
\ee
with $C = \frac{1}{2}$ and $N_c=3$,  the $Q_2$ operator can be expressed as

\bea
Q_2 = \frac{1}{N_c} \bar{u}_{\alpha} \gamma^{\mu} (1-\gamma_5) q_{\alpha} \bar{b}_{\beta} \gamma_{\mu} (1-\gamma_5) c_{\beta} + \frac{1}{C} \bar{u}_{\alpha} (T_a)_{\alpha \beta} \gamma^{\mu} (1-\gamma_5) q_{\beta} \bar{b}_{\rho} (T_a)_{\rho \lambda} \gamma_{\mu} (1-\gamma_5) c_{\lambda} \, .
\eea
The effective Hamiltonian is rewritten as
\be
\mathcal{H}_{eff}= \frac{G_F}{\sqrt{2}} V_{c b}^* V_{u q}\left(C_0 (\mu) Q_0 (\mu) + C_8 (\mu) Q_8 (\mu) \right)  \, , \label{Hamilnew}
\ee
in terms of the operators
\bea
Q_0 &=& \bar{u}_{\alpha} \gamma^{\mu} (1-\gamma_5) q_{\alpha} \bar{b}_{\beta} \gamma_{\mu} (1-\gamma_5) c_{\beta}  \nn \\
Q_8 &=& \bar{u}_{\alpha} (T_a)_{\alpha \beta} \gamma^{\mu} (1-\gamma_5) q_{\beta} \bar{b}_{\rho} (T_a)_{\rho \lambda} \gamma_{\mu} (1-\gamma_5) c_{\lambda} .
\eea
$Q_0$ and $Q_8$ are color-singlet and  color-octet operators contributing to the decay.
The Wilson coefficients in \eqref{Hamilnew}  are related to the ones in \eqref{Hamil}:
\be
C_0 = C_1 + \frac{1}{N_c} C_2= a_1\,\, , \qq C_8 = \frac{1}{C} C_2 .
\ee

We consider the processes  $B_c^+ \to c \bar{c}(P) \, M_{P (V)}$, where  $c \bar{c}(P)$ is one of the P-wave charmonium, and  $P$,$V$  a light pseudoscalar or a vector meson. The decay width
\bea
\Gamma (B_c^+ \to c \bar{c}(P) \, M_{P (V)}) = \frac{\lvert \vec q \rvert}{8 \pi m_{B_c}^2} \lvert \mathcal{A}(B_c^+ \to c \bar{c}(P) \, M_{P (V)})\rvert ^2,
\label{eq:gamma}
\eea
with  $\lvert \vec q \rvert = \frac{1}{2 m_{B_c}} \sqrt{\lambda(m_{B_c}^2,m_{c \bar{c}}^2,m_{P(V)}^2)} $ the  three-momentum of one of the final mesons in the $B_c$ rest-frame and  $\lambda(x,y,z) = x^2 + y^2 + z^2 - 2xy -2xz -2yz$ the K\"all\'en function, involves the matrix element 

\bea
\mathcal{A}(B_c^+ \to c \bar{c}(P) \, M_{P (V)}) = \bra{c\bar{c}(P) \, M_{P(V)}} \mathcal{H}_{eff} \ket{B_c^+}.
\label{eq:ampl1}
\eea
In naive factorization the contribution of the color-octet operator is discarded and  the matrix element is factorized as 

\bea
\mathcal{A}(B_c^+ \to c \bar{c}(P) \, M_{P (V)}) = \frac{G_F}{\sqrt{2}} V_{c b}^* V_{uq} a_1 (\mu) \bra{c \bar{c}(P)}\bar{b} \gamma_{\mu} (1-\gamma_5) c \ket{B_c^+} \bra{M_{P (V)}}\bar{u} \gamma^{\mu} (1-\gamma_5) q \ket{0}. \nn \\
\label{eq:ampl}
\eea
The parametrization is used for the current-particle matrix elements,
\bea
\bra{M_P(q)}\bar{u} \gamma^{\mu} (1 &-& \gamma_5) q \ket{0} = - i f_{P} q^{\mu} \label{AnnMat}\\ \bra{M_V(q,\epsilon_V)}\bar{u} \gamma^{\mu} (1 &-& \gamma_5) q \ket{0} = m_V f_{V} \epsilon^{*\mu}_V  \, , \label{AnnMatVec} 
\eea
and for the matrix elements in terms of form factors:
\bea
\langle S(p')|\bar{b}\gamma^\mu \gamma_5 c|{B_c} (p)\rangle
&=& f_0^{S}(q^2) \frac{m_{B_c}^2 - m_{S}^2}{q^2} q^{\mu} + f_+^{S}(q^2) \left( p^{\mu} + p'^{\mu} - \frac{m_{B_c}^2 - m_{S}^2}{q^2} q^{\mu} \right), \\
\langle A(p', \epsilon)|\bar{b}\gamma^\mu c|{B_c} (p)\rangle
&=& -i \Big[2 m_{A} A_0^{A}(q^2) \frac{\epsilon^* \cdot q}{q^2} q^{\mu} + (m_{B_c} + m_{A}) A_1^{A}(q^2) \left(\epsilon^{* \mu}-\frac{\epsilon^* \cdot q}{q^2} q^{\mu} \right) \nn \\
&-& A_2^{A}(q^2) \frac{\epsilon^* \cdot q}{m_{B_c}+m_{A}} \left(p^{\mu} + p'^{\mu}-\frac{m_{B_c}^2-m_{A}^2}{q^2} q^{\mu}\right) \Big],\\
\langle A(p', \epsilon)|\bar{b}\gamma^\mu \gamma_5 c|{B_c} (p)\rangle
&=& \frac{2 V^{A}(q^2)}{m_{B_c} + m_{A}}\varepsilon^{\mu\nu\rho\sigma} \epsilon_{\nu}^* p_{\rho}' p_{\sigma}, \\
\langle T(p', \eta)|\bar{b}\gamma^\mu c|{B_c} (p)\rangle &=& \frac{2 V^{T}(q^2)}{m_{B_c} \left(m_{B_c} + m_{T}\right)} \varepsilon^{\mu\nu\rho\sigma} \eta_{\nu \alpha}^* p_{\rho}' p_{\sigma} q^{\alpha}, \\
\langle T(p', \eta)|\bar{b}\gamma^\mu \gamma_5 c|{B_c} (p)\rangle
&=& \Big[2 m_{T} A_0^{T}(q^2) \frac{\eta^{* \alpha \beta} q_{\beta}}{q^2} q^{\mu} + (m_{B_c} + m_{T}) A_1^{T}(q^2) \left(\eta^{* \mu \alpha}-\frac{\eta^{* \alpha \beta} q_{\beta}}{q^2} q^{\mu}\right) \nn \\ 
&-& A_2^{T}(q^2)\frac{\eta^{* \alpha \beta} q_{\beta}}{m_{B_c} + m_{T}} \left(p^{\mu} + p'^{\mu} -\frac{m_{B_c}^2-m_{T}^2}{q^2} q^{\mu}\right) \Big] \frac{-i q_{\alpha}}{m_{B_c}} , 
\eea
\noindent
with $q=p-p^\prime$, and $S=\chi_{c0}$, $A=\chi_{c1}, h_c$ and $T=\chi_{c2}$. $\epsilon_V$, $\epsilon$ and $\eta$ are the polarization vector and tensor of the light vector meson,  and of the charmonium axial-vector and tensor state, respectively. The conditions are imposed
\bea
f_0^S(0)&=&f_+^S(0), \\
A_0^{A(T)}(0)&=& \frac{m_{B_c}+m_{A(T)}}{2 m_{A(T)}}A_1^{A(T)}(0)-\frac{m_{B_c}-m_{A(T)}}{2 m_{A(T)}}A_2^{A(T)}(0).
\eea
\noindent 
The light meson selects the form factors contributing to the decay width: $f_0$ and $A_0^{A(T)}$ at $q^2 = m_{P}^2$ for a light pseudoscalar, all the others at $q^2 = m_{V}^2$ for a light vector meson.
For the transitions to a pseudoscalar light meson we have, using factorization:

\bea
\Gamma (B_c^+ \to \chi_{c0} \, M_{P}) &=&  \Lambda \left[f_0(m_P ^2)\right]^2 f_{P}^2 \lvert V_{cb} \rvert ^2 \lvert V_{uq} \rvert ^2 \left(m_{B_c}^2 -m_{\chi_{c0}}^2 \right)^2  \lambda^{\frac{1}{2}}(m_{B_c}^2,m_{\chi_{c0}} ^2,m_{P}^2) \nn\\ \label{eq:GammaChic0P}\\
\Gamma (B_c^+ \to A \, M_{P}) &=& \Lambda \left[A_0^A(m_P^2)\right]^2 f_{P}^2 \lvert V_{cb} \rvert ^2 \lvert V_{uq} \rvert ^2  \lambda^{\frac{3}{2}}(m_{B_c}^2,m_{A} ^2,m_{P}^2)\label{eq:GammaAP}\\
\Gamma (B_c^+ \to \chi_{c2} \, M_{P}) &=& \Lambda \frac{\left[A^{\chi_{c2}}_0(m_P ^2)\right]^2 f_{P}^2 \lvert V_{cb} \rvert ^2 \lvert V_{uq} \rvert ^2 \lambda^{\frac{5}{2}}(m_{B_c}^2,m_{\chi_{c2}}^2,m_{P}^2)}{6 m_{B_c}^2 m_{\chi_{c2}}^2}. \label{eq:GammaChic2P}
\eea
\noindent
where
\bea
\Lambda = \frac{a_1(m_b)^2 G_F^2}{32 \, \pi \, m_{B_c}^3}.
\label{eq:lambda}
\eea

In case of a light vector meson we have:

\bea
\Gamma (B_c^+ \to \chi_{c0} \, M_{V}) &=&  \Lambda \left[f_+^{\chi_{c0}}(m_{V}^2)\right]^2 f_{V}^2 \lvert V_{cb} \rvert ^2 \lvert V_{uq} \rvert ^2 \lambda^{\frac{3}{2}}(m_{B_c}^2,m_{\chi_{c0}}^2,m_{V}^2) \label{eq:GammaChic0V}\\
\Gamma (B_c^+ \to A \, M_{V}) &=& \Lambda \frac{f_{V}^2 \lvert 
V_{cb} \rvert ^2 \lvert V_{uq} \rvert ^2 \lambda^{\frac{1}{2}}(m_{B_c}^2,m_{A} 
^2,m_{V}^2)}{{4 
m_{A}^2 (m_{B_c}+m_{A})^2}}\bigg\{\left[A_1^{A}(m_{V}^2)\right]^2 \left( m_{B_c}+m_{A}\right)^4 \nn \\ 
& \times& \left( \lambda(m_{B_c}^2,m_{A}^2,m_{V}^2) + 12 m_{A}^2 m_{V}
^2\right)+ \left[A_2^{A}(m_{V}^2)\right]^2 \lambda^2(m_{B_c}^2,m_{A}^2,m_{V}
^2) \\ \nn
&+& 2 A_1^{A}(m_{V}^2) A_2^{A}(m_{V}^2) \lambda(m_{B_c}
^2,m_{A}^2,m_{V}^2) \left( m_{B_c}+m_{A}\right)^2\left( m_{B_c}
^2-m_{A}^2-m_{V}^2\right)\\ \nn
&+& 8 \left[V^{A}(m_{V}^2)\right]^2 m_{A}^2 
m_{V}^2 \lambda(m_{B_c}^2,m_{A}^2,m_{V}^2) \bigg\}\label{eq:GammaAV}\\
\Gamma (B_c^+ \to \chi_{c2} \, M_{V}) &=& \Lambda \frac{f_{V}^2 \lvert 
V_{cb} \rvert ^2 \lvert V_{uq} \rvert ^2 \lambda^{\frac{3}{2}}(m_{B_c}^2,m_{\chi_{c2}} 
^2,m_{V}^2)}{{24 m_{B_c}^2 
m_{\chi_{c2}}^4 (m_{B_c}+m_{\chi_{c2}})^2}}\bigg\{\left[A_1^{\chi_{c2}}(m_{V}^2)\right]^2 ( m_{B_c}+m_{\chi_{c2}})^4 \nn \\ 
& \times & \left( \lambda(m_{B_c}^2,m_{\chi_{c2}}^2,m_{V}^2) + 10 m_{\chi_{c2}}^2 m_{V}
^2\right) + \left[A_2^{\chi_{c2}}(m_{V}^2)\right]^2 \lambda^2(m_{B_c}^2,m_{\chi_{c2}}^2,m_{V}
^2)\nn \\
&-& 2 A_1^{\chi_{c2}}(m_{V}^2) A_2^{\chi_{c2}}(m_{V}^2) \left( m_{B_c}+m_{\chi_{c2}}\right)^2\big[ m_{B_c}
^6-3 m_{B_c}^4 \left(m_{\chi_{c2}}^2+m_{V}^2\right) \nn \\ 
&-& \left( m_{\chi_{c2}}^2-m_{V}^2\right)^2 \left( m_{\chi_{c2}}^2 + m_{V}^2 \right) + m_{B_c}^2 \left( 3 m_{\chi_{c2}}^4 + 2 m_{\chi_{c2}}^2 m_{V}^2 + 3 m_{V}^4 \right) \big] \nn \\ 
&+& 6 \left[V^{\chi_{c2}}(m_{V}^2)\right]^2 m_{\chi_{c2}}^2 
m_{V}^2 \lambda(m_{B_c}^2,m_{\chi_{c2}}^2,m_{V}^2) \bigg\}, \label{eq:GammaChic2V}
\eea
\noindent
The Wilson coefficient $a_1(\mu)$ is evaluated at the scale $m_b$. $a_1(\mu)$ is a combination of the  coefficients $C_1(\mu)$ and $C_2(\mu)$, whose two-loop computation is described in \cite{Buras:1998raa}. The naive factorization approach does not take into account nonfactorizable effects, e.g. the final-state interaction, the gluon exchanges between quarks of final mesons, which are expected to be sizable if both the final hadrons are heavy and the phase-space is small. On the basis of the Bjorken's colour transparency argument \cite{Bjorken:1988kk}, if the mesons do not have enough energy to escape from the colour field of each other,  nonfactorizable interactions emerge. It is possible to partially encode such effects replacing the coefficient  $a_1(\mu)$  ($a_2(\mu)$ for colour suppressed processes),  by the  effective coefficient $a_1^{eff}(\mu)$ ($a_2^{eff}(\mu)$), treated as a phenomenological parameter \cite{Neubert:1997uc,Cheng:1998kd}. Since we are considering processes involving one  light final meson, we  neglect the nonfactorizable terms and use $a_1(m_b) = 1.025$. A recent use of factorization for nonleptonic decays of $B$ and $D$ mesons can be found in \cite{Yu:2022ngu}.
Factorization and nonfactorizable contributions in  nonleptonic  $B$ meson decays, and their effects in the heavy-quark limit, are reviewed in \cite{Beneke:2000ry}.
For $B_c$ decays such terms have been considered in the PQCD approach (in \cite{Xiao:2013lia,Rui:2017pre,Liu:2018kuo} and  in the references quoted therein) and in the QCD factorization \cite{Ebert:2010zu,Dhir:2012sv,Sun:2015exa}.

\section{Observables}\label{Observables}

The hadronic form factors of $B_c$ to the $P$–wave charmonium 4-plet $\chi_{cJ}, h_c$ can be expressed near the zero-recoil point in terms of universal functions, performing an expansion in QCD in the relative velocity of the heavy quarks and in $1/m_Q$  \cite{Colangelo:2022awx}. At the leading order in the expansion a single universal function $\Xi(q^2)$ parametrizes all such form factors. The following results are based on the extrapolation of the universal form factor from the zero-recoil point to the $q^2$ value corresponding to mass of the final light meson. Even though the next to leading order (NLO) in the expansion is presented in the analysis \cite{Colangelo:2022awx}, the number of universal functions parametrizing these contributions does not allow us to construct useful observables. Therefore, the NLO is ignored and the analysis is performed considering a single universal function.

The relations for the form factors relevant for the transition to a light pseudoscalar meson are:
\bea
f_0^{\chi_{c0}}(q^2) &=&- \frac{((m_{B_c}-m_{\chi_{c0}})^2-q^2)((m_{B_c}+m_{\chi_{c0}})^2 -q^2)}{4 \sqrt{3} (m_{B_c}-m_{\chi_{c0}})(m_{B_c}m_{\chi_{c0}})^{3/2}} \Xi(q^2), \label{FFchic0}\\
A_0^{\chi_{c1}}(q^2) &=& 0, \label{FFchic1}\\
A_0^{h_c}(q^2) &=& - i\frac{(m_{B_c}-m_{h_c})((m_{B_c}+ m_{h_c})^2 - q^2)}{4 (m_{B_c} m_{h_c})^{3/2}} \Xi(q^2), \label{FFhc} \\
A_0^{\chi_{c2}}(q^2) &=& i \frac{m_{B_c} + m_{\chi_{c2}}}{2 \sqrt{m_{B_c} m_{\chi_{c2}}}} \Xi(q^2), \label{FFchic2}
\eea
The relations for the form factors relevant for the decays to a light vector meson are:

\bea
f_+^{\chi_{c0}}(q^2) &=&- \frac{((m_{B_c}+m_{\chi_{c0}})^2-q^2)(m_{B_c}-m_{\chi_{c0}})}{4 \sqrt{3} (m_{B_c}m_{\chi_{c0}})^{3/2}} \Xi(q^2), \label{FFForV1}\\
V^{\chi_{c1}}(q^2) &=& - \frac{((m_{B_c}+m_{\chi_{c1}})^2-q^2)(m_{B_c}+m_{\chi_{c1}})}{4 \sqrt{2} (m_{B_c}m_{\chi_{c1}})^{3/2}} \Xi(q^2), \label{FFForV2}\\
A_1^{\chi_{c1}}(q^2) &=& - \frac{m_{B_c}^4+(m_{\chi_{c1}}^2-q^2)^2-2 m_{B_c}^2 (m_{\chi_{c1}}^2+q^2)}{4 \sqrt{2} (m_{B_c}m_{\chi_{c1}})^{3/2}(m_{B_c} +m_{\chi_{c1}})} \Xi(q^2), \label{FFForV3} \\
A_2^{\chi_{c1}}(q^2) &=& \frac{(m_{B_c}^2-m_{\chi_{c1}}^2-q^2)(m_{B_c}+m_{\chi_{c1}})}{4 \sqrt{2} (m_{B_c}m_{\chi_{c1}})^{3/2}} \Xi(q^2), \\
V^{\chi_{c2}}(q^2) &=&  \frac{m_{B_c}+m_{\chi_{c2}}}{2 \sqrt{m_{B_c}m_{\chi_{c2}}}} \Xi(q^2),\label{FFForV4}\\
A_1^{\chi_{c2}}(q^2) &=& i \frac{((m_{B_c}+m_{\chi_{c2}})^2-q^2)}{2  \sqrt{m_{B_c}m_{\chi_{c2}}}(m_{B_c} +m_{\chi_{c2}})} \Xi(q^2), \label{FFForV5}\\
A_2^{\chi_{c2}}(q^2) &=& i \frac{m_{B_c}+m_{\chi_{c2}}}{2 \sqrt{m_{B_c}m_{\chi_{c2}}}} \Xi(q^2),\label{FFForV6} \\
V^{h_c}(q^2) &=& 0 \label{FFForV7}, \\
A_1^{h_{c}}(q^2) &=& 0, \label{FFForV8} \\
A_2^{h_{c}}(q^2) &=&i \frac{m_{h_c} (m_{B_c}+m_{h_{c}})^2}{2 (m_{B_c}m_{h_{c}})^{3/2}} \Xi(q^2). \label{FFForV9}
\eea
Remarkably, at LO some form factors vanish: a consequence of 
 Eq.~\eqref{FFchic1} is that the $B_c$  transition to $\chi_{c1}$  and a light pseudoscalar  is suppressed.
 \noindent
 
We use the relations (\ref{FFchic0})-(\ref{FFForV9}) and the parameters  in Table~\ref{tab:input}. The decay constants $f_{\rho}$ and $f_{K^*}$ are obtained from  $\mathcal{B}(\tau^- \to \rho^- \nu_{\tau}) = (25.19 \pm 0.3) \times 10^{-2}$ and $\mathcal{B}(\tau^- \to K^{*-} \nu_{\tau}) = (1.20 \pm 0.07) \times 10^{-2} $ \cite{Workman:2022ynf}. For the mass of the not yet discovered $h_c(2P)$  there are predictions from different models, a few results  are presented in Table~\ref{tab:input}. We use  $m_{h_c(2P)} = 3.902 \, \gev$ obtained in  \cite{Zhou:2017dwj}, since in such a model the masses of the $\chi_{cJ}(2P)$ triplet agree with measurements.
 
\begin{table}[!h]
\center{\begin{tabular}{|l|l|}
\hline
$G_F = 1.16637(1)\times 10^{-5}\gev^{-2}$\hfill 	& 
$\lvert V_{cb} \rvert = 0.04182^{+0.00085}_{-0.00074}$\hfill \\
$m_{B_c}= 6.274 \, \gev$\hfill						
&	$\lvert V_{ud} \rvert = 0.97435 \pm 0.00016$\hfill \\
$\tau_{B_c} = 0.510 \pm 0.009 \, \textrm{ps}$				& 	
$\lvert V_{us} \rvert = 0.22500 \pm 0.00067$\hfill \\\hline
$m_{B_s^0} = 5.367 \, \gev$ \hfill						
& $\lvert V_{cs} \rvert = 0.97349 \pm 0.00016$ \hfill \\
$m_{J/\psi} = 3.097 \,\gev$ \hfill & 
\\ \hline
$m_{h_c(1P)} = 3.525 \, \gev$\hfill   & $m_{h_c(2P)} = 
\begin{cases}
3.897 \, \gev \text{\cite{Azizi:2017izn}} \\
3.902 \, \gev \text{\cite{Zhou:2017dwj}}\\
3.934 \, \gev \text{\cite{Barnes:2005pb}}\\
3.956 \, \gev \text{\cite{Godfrey:1985xj}}
\end{cases}$ \hfill \\ 
$m_{\chi_{c0}(1P)} = 3.415 \, \gev$\hfill  & $m_{\chi_{c0}(2P)} = 3.860 \, \gev$ \hfill \\ 
$m_{\chi_{c1}(1P)} = 3.511 \, \gev$ \hfill & 
$m_{\chi_{c1}(2P)} = 3.872 \, \gev$\hfill  \\ $m_{\chi_{c2}(1P)} = 3.556 \, \gev$ \hfill & 
$m_{\chi_{c2}(2P)} = 3.930 \, \gev$ \hfill \\ 
\hline
$m_{\pi} = 139.57 \, \mev$\hfill   & $f_{\pi} = 130.2\, \mev$ \hfill \\ 
$m_{K} = 493.677 \, \mev$\hfill  & $f_{K} = 155.7 \, \mev$ \hfill \\ 
$m_{\rho} = 770 \, \mev$ \hfill & 
$f_{\rho} = 209 \, \mev$\hfill  \\ $m_{K^*} = 892 \, \mev$ \hfill & 
$f_{K^*} = 205 \, \mev$ \hfill \\ 
\hline
\end{tabular}  }
\caption {\small Input parameters  \cite{Workman:2022ynf}.}
\label{tab:input}
\end{table}
\noindent

\subsection*{$\pi^+$ and $K^+$ modes}
Using Eqs.~\eqref{FFchic0}-\eqref{FFchic2} in Eqs.~\eqref{eq:GammaChic0P}-\eqref{eq:GammaChic2P} we have for  the $B_c$ transition to  $\pi^+$:

\bea
&&\Gamma (B_c^+ \to \chi_{c0} \pi^+) = \Lambda_{\pi} \frac{(m_{B_c}+m_{\chi_{c0}})^2 \lambda^{\frac{5}{2}}(m_{B_c}^2,m_{\chi_{c0}}^2,m_{\pi}^2)}{3 m_{\chi_{c0}}^3}\, 
\left[\Xi(m_{\pi}^2)\right]^2, \label{eq:GammaChic0Pi}\\
&&\Gamma (B_c^+ \to \chi_{c2} \pi^+) = 2 \Lambda_{\pi} \frac{(m_{B_c}+m_{\chi_{c2}})^2 \lambda^{\frac{5}{2}}(m_{B_c}^2,m_{\chi_{c2}}^2,m_{\pi}^2)}{3 m_{\chi_{c2}}^3}\, 
\left[\Xi(m_{\pi}^2)\right]^2 , \\
&&\Gamma (B_c^+ \to h_{c} \pi^+) = \Lambda_{\pi} \frac{(m_{B_c}-m_{h_{c}})^2 ((m_{B_c}+m_{h_c})^2-m_{\pi}^2)^2 \lambda^{\frac{3}{2}}(m_{B_c}^2,m_{h_c}^2,m_{\pi}^2)}{m_{h_c}^3} \left[\Xi(m_{\pi}^2)\right]^2 \label{eq:GammahcPi}
\eea
\noindent 
with $\Lambda_{\pi} = \Lambda \frac{ f_{\pi}^2 \lvert V_{cb} \rvert ^2 \lvert V_{ud} \rvert ^2}{16 m_{B_c}^3}$,  where $\Lambda$ is given in Eq.~\eqref{eq:lambda}.
From the measurement $\mathcal{B} (B_c^+ \to \chi_{c0} \pi^+) = (2.4^{+0.9}_{-0.8}) \times 10^{-5}$ and from Eq.~\eqref{eq:GammaChic0Pi} the value $\Xi(m_{\pi}^2) = 0.243^{+0.045}_{-0.040}$ is obtained, allowing us to predict the branching fractions to  $\chi_{c2}(1P)$ and $h_c(1P)$:

\bea
\mathcal{B}(B_c^+ \to \chi_{c2} \pi^+)=3.65^{+1.37}_{-1.22}\times 10^{-5}, \qquad \mathcal{B}(B_c^+ \to h_c \pi^+)=5.83^{+2.19}_{-1.94}\times 10^{-5}.
\eea
For the $2P$ excitations there are no measurements that can be exploited to determine the  universal form factor. However, since  this quantity  cancels in the ratios of different channels involving charmonia belonging to the same 4-plet, the predictions in Table~\ref{tab:piK} can be derived.


\begin{table}[!t]
\begin{adjustbox}{width = 0.6\textwidth,center}
\begin{tabular}{c|ccc}

  & $\frac{\mathcal{B}(B_c^+ \to \chi_{c0} \, \pi^+)}{\mathcal{B}(B_c^+ \to \chi_{c2} \, \pi^+)}$  & $\frac{\mathcal{B}(B_c^+ \to h_c \, \pi^+)}{\mathcal{B}(B_c^+ \to \chi_{c0} \, \pi^+)}$ & $\frac{\mathcal{B}(B_c^+ \to h_c \, \pi^+)}{\mathcal{B}(B_c^+ \to \chi_{c2} \,\pi^+)}$ \\
\hline
$1P$ & $0.658$ & $2.429$ & $1.597$ \\
$2P$ & $0.583$ & $2.746$ & $1.601$  \\
\hline \hline
 & $\frac{\mathcal{B}(B_c^+ \to \chi_{c0} \, K^+)}{\mathcal{B}(B_c^+ \to \chi_{c2} \, K^+)}$  & $\frac{\mathcal{B}(B_c^+ \to h_c \, K^+)}{\mathcal{B}(B_c^+ \to \chi_{c0} \, K^+)}$ & $\frac{\mathcal{B}(B_c^+ \to h_c \, K^+)}{\mathcal{B}(B_c^+ \to \chi_{c2} \, K^+)}$ \\
\hline
$1P$ & $0.663$ & $2.482$ & $1.645$ \\
$2P$ & $0.586$ & $2.845$ & $1.668$  
\end{tabular}
\end{adjustbox}
\caption {\small Ratios of branching fractions of $B_c$ decays in charmonium state and a $\pi^+$ and a $K^+$ meson.}
\label{tab:piK}
\end{table}
\noindent

The modes with $\chi_{c1}(1P)$ and $\chi_{c1}(2P)$  are suppressed: the observation of such a suppression for $\chi_{c1}(3872)$ would be in favour of the identification of $X(3872)$ as an ordinary charmonium state.

For the branching ratios with $K^+$ in the finale state the predictions are also in Table~\ref{tab:piK}.

\subsection*{$\rho^+$ and $K^{*+}$ modes}
Exploiting Eqs.~\eqref{eq:GammaChic0V}-\eqref{eq:GammaChic2V} and Eqs.~\eqref{FFForV1}-\eqref{FFForV9} we obtain the results in Table \ref{tab:rho} for the $\rho^+$ modes, and in Table~\ref{tab:Kstar} for the $K^{*+}$ modes.

\begin{table}[!b]
\begin{adjustbox}{width = 0.6\textwidth,center}
\begin{tabular}{c|ccc}

  & $\frac{\mathcal{B}(B_c^+ \to \chi_{c1}\rho^+)}{\mathcal{B}(B_c^+ \to \chi_{c0} \rho^+)}$  & $\frac{\mathcal{B}(B_c^+ \to \chi_{c1} \rho^+)}{\mathcal{B}(B_c^+ \to \chi_{c2} \rho^+)}$ & $\frac{\mathcal{B}(B_c^+ \to \chi_{c0}\rho^+)}{\mathcal{B}(B_c^+ \to \chi_{c2} \rho^+)}$ \\
\hline
$1P$ & $0.206$ & $0.122$ & $0.590$ \\
$2P$ & $0.315$ & $0.159$ & $0.503$ \\
\hline \hline
& $\frac{\mathcal{B}(B_c^+ \to h_c \rho^+)}{\mathcal{B}(B_c^+ \to \chi_{c0} \rho^+)}$  & $\frac{\mathcal{B}(B_c^+ \to h_c \rho^+)}{\mathcal{B}(B_c^+ \to \chi_{c1} \rho^+)}$ & $\frac{\mathcal{B}(B_c^+ \to h_c \rho^+)}{\mathcal{B}(B_c^+ \to \chi_{c2} \rho^+)}$ \\
\hline
$1P$ & $2.226$ & $10.790$ & $1.312$ \\
$2P$ & $2.449$ & $7.770$ & $1.232$ \\
\end{tabular}
\end{adjustbox}
\caption {\small Ratios of branching fractions of $B_c$ decays in charmonium state and a $\rho^+$ meson.}
\label{tab:rho}
\end{table}
\noindent

%


\begin{table}[!h]
\begin{adjustbox}{width = 0.6\textwidth,center}
\begin{tabular}{c|ccc}

  & $\frac{\mathcal{B}(B_c^+ \to \chi_{c1}K^{*+})}{\mathcal{B}(B_c^+ \to \chi_{c0} K^{*+})}$  & $\frac{\mathcal{B}(B_c^+ \to \chi_{c1} K^{*+})}{\mathcal{B}(B_c^+ \to \chi_{c2} K^{*+})}$ & $\frac{\mathcal{B}(B_c^+ \to \chi_{c0}K^{*+})}{\mathcal{B}(B_c^+ \to \chi_{c2} K^{*+})}$ \\
\hline
$1P$ & $0.276$ & $0.157$ & $0.570$ \\
$2P$ & $0.422$ & $0.203$ & $0.481$ \\
\hline \hline
& $\frac{\mathcal{B}(B_c^+ \to h_c K^{*+})}{\mathcal{B}(B_c^+ \to \chi_{c0} K^{*+})}$  & $\frac{\mathcal{B}(B_c^+ \to h_c K^{*+})}{\mathcal{B}(B_c^+ \to \chi_{c1} K^{*+})}$ & $\frac{\mathcal{B}(B_c^+ \to h_c K^{*+})}{\mathcal{B}(B_c^+ \to \chi_{c2} K^{*+})}$ \\
\hline
$1P$ & $2.159$ & $7.834$ & $1.231$ \\
$2P$ & $2.350$ & $5.568$ & $1.131$ \\
\end{tabular}
\end{adjustbox}
\caption {\small Ratios of branching fractions of $B_c$ decays in charmonium state and $K^{*+}$ meson.}
\label{tab:Kstar}
\end{table}

\begin{figure}[!b]
\begin{center}
\includegraphics[width = 0.4\textwidth]{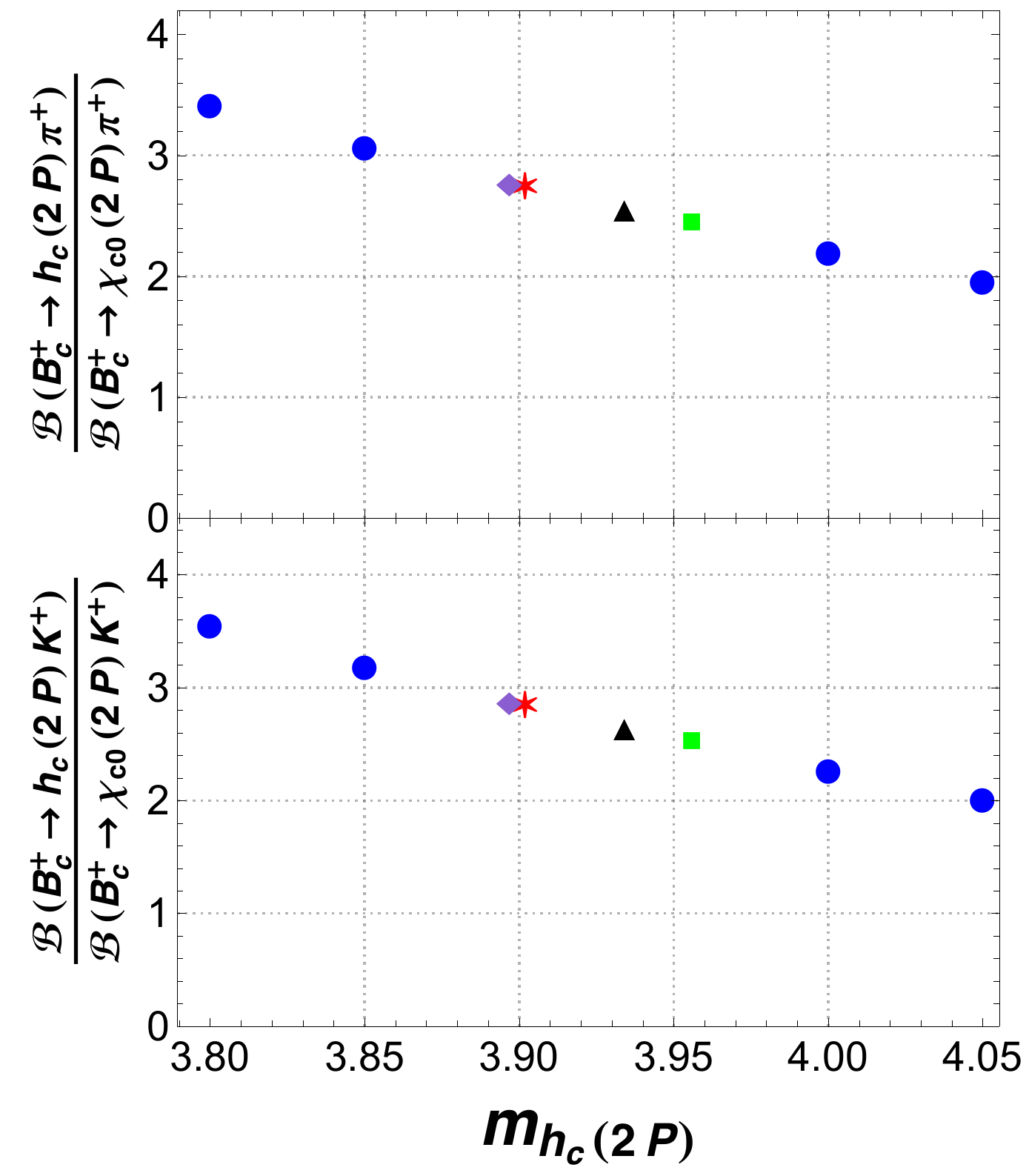}\hskip 0.2cm \includegraphics[width = 0.4\textwidth]{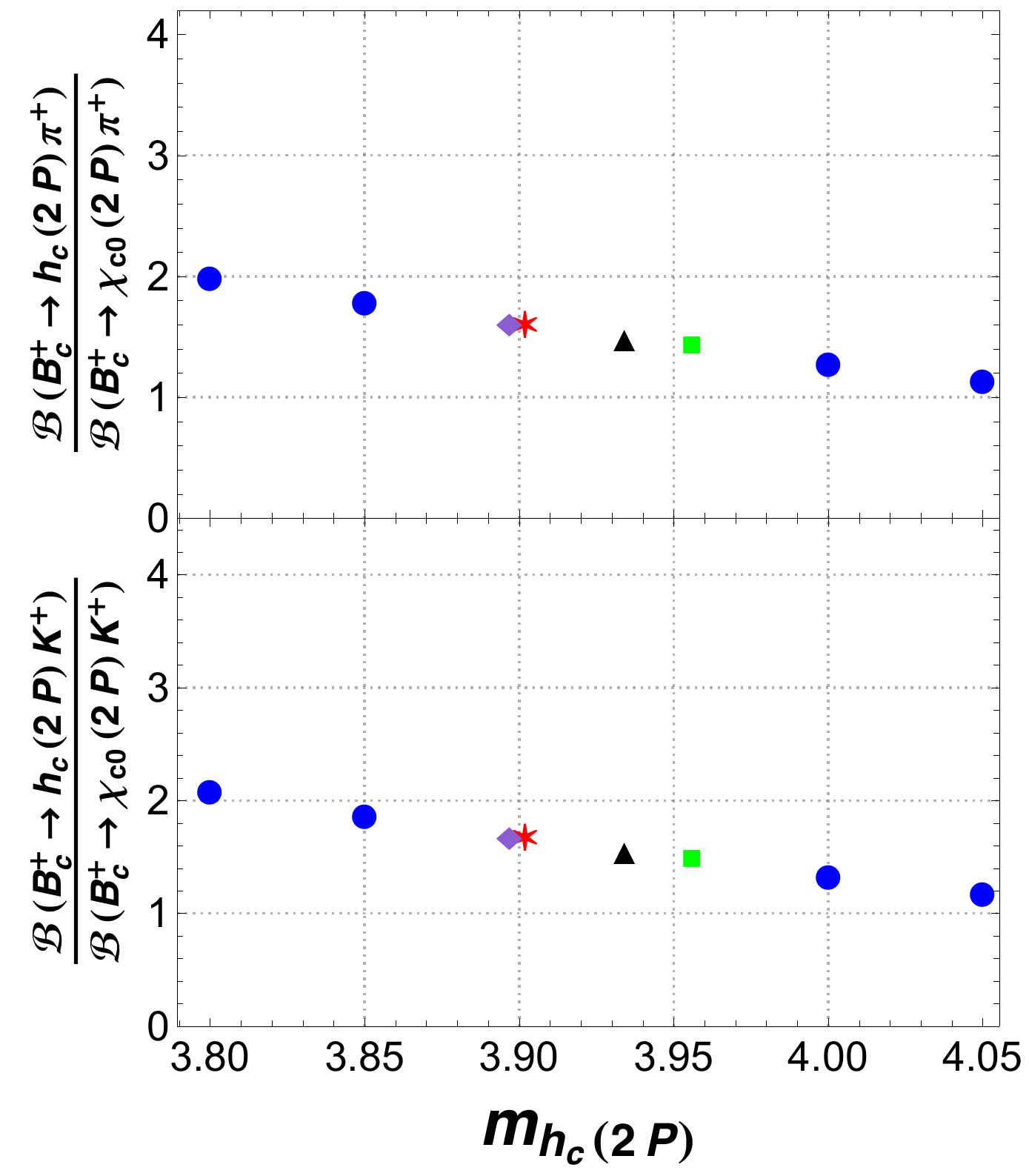}
    \caption{\small Ratios of  branching fractions ${\cal B}(B_c^+ \to h_c(2P)\, \pi^+)/{\cal B}(B_c^+ \to \chi_{c0}(2P) \, \pi^+)$ (first row, left panel) and ${\cal B}(B_c^+ \to h_c(2P)\, \pi^+)/{\cal B}(B_c^+ \to \chi_{c2}(2P) \, \pi^+)$ (first row, right panel) versus  $m_{h_c(2P)}$. The plots in the second row refer to  $K^+$ in the final state. The red stars correspond to the $h_c(2P)$ mass evaluated in \cite{Zhou:2017dwj} and used in the present study,  the violet diamonds, the black triangles and the green squares to $m_{h_c(2P)}$ computed in \cite{Azizi:2017izn}, \cite{Barnes:2005pb} and \cite{Godfrey:1985xj}, respectively.} \label{Fig1}
\end{center}
\end{figure}

\begin{figure}[!h]
\begin{center}
\includegraphics[width = 0.32\textwidth]{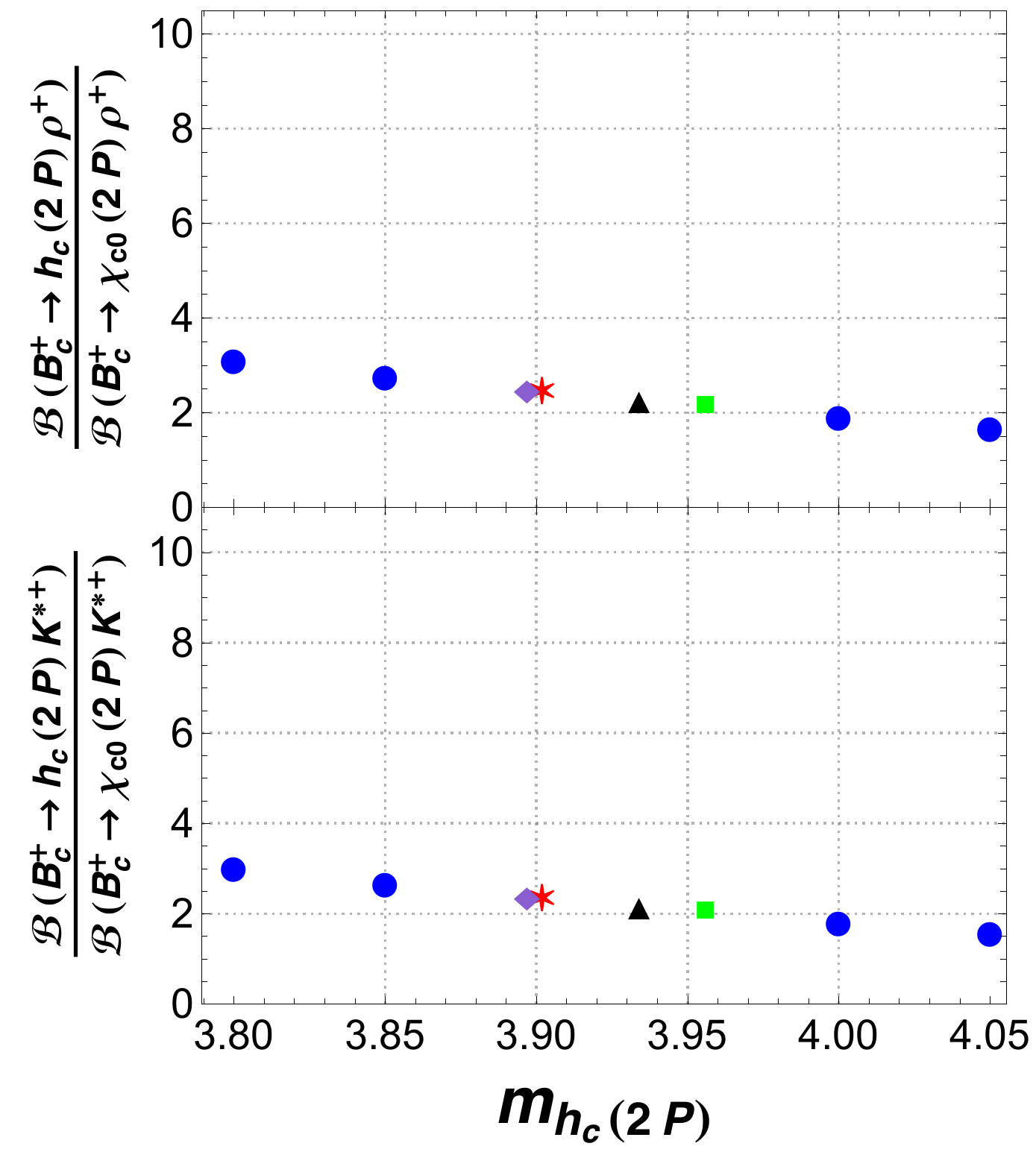} \includegraphics[width = 0.32\textwidth]{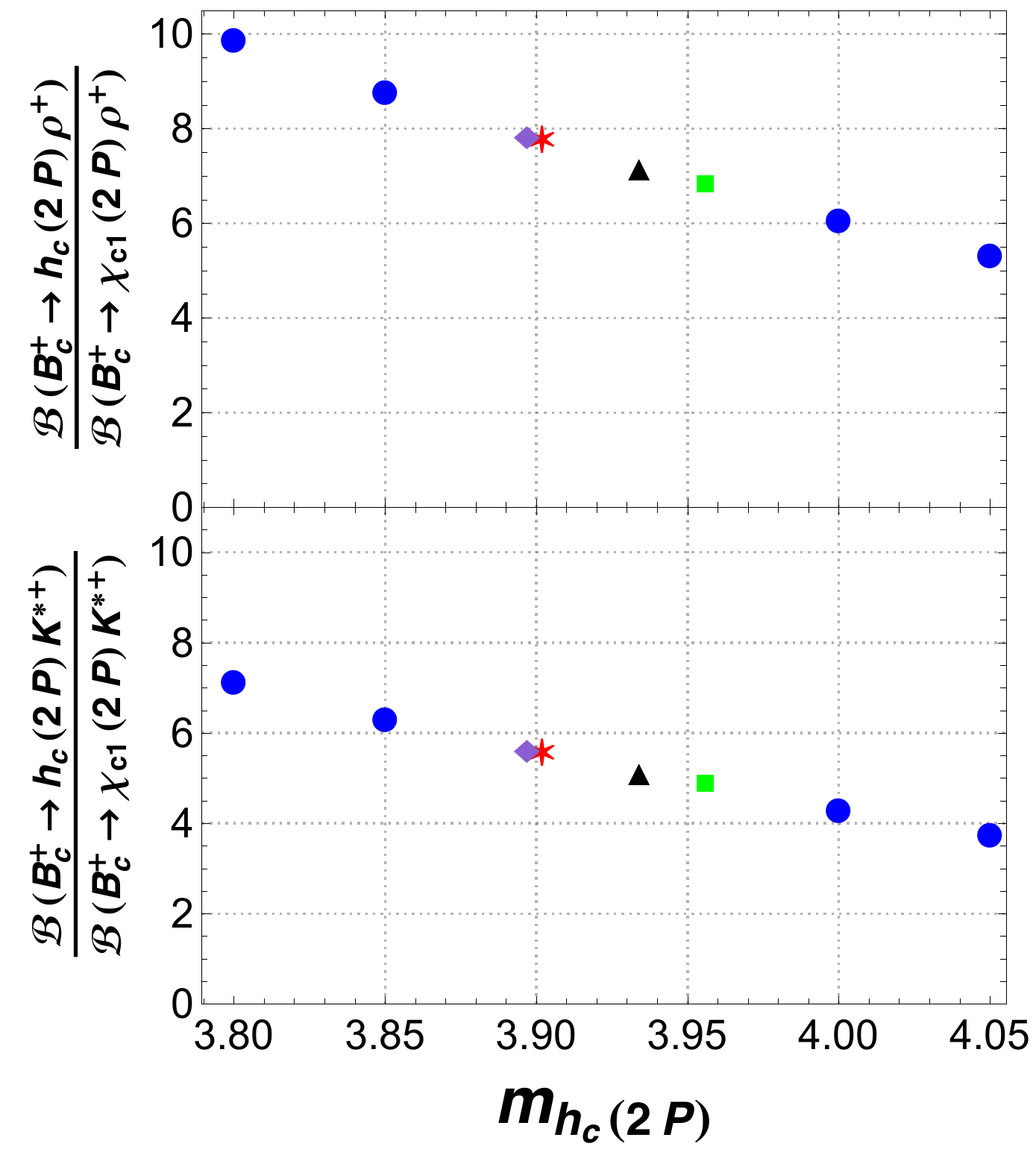}
\includegraphics[width = 0.32\textwidth]{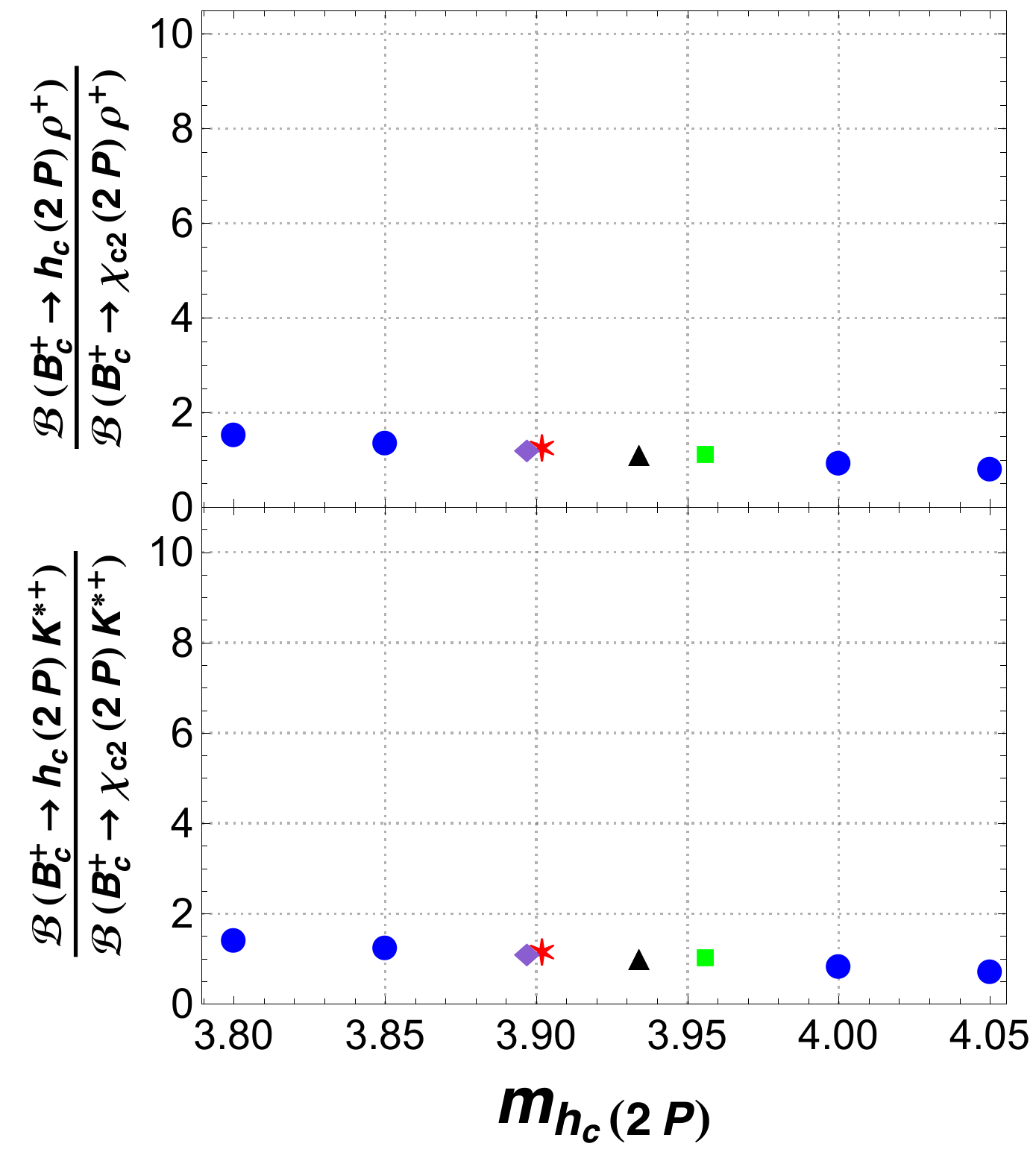}
    \caption{\small Ratios of  branching fractions  ${\cal B}(B_c^+ \to h_c(2P)\, \rho^+)/{\cal B}(B_c^+ \to \chi_{c0}(2P) \, \rho^+)$ (first row, left panel), ${\cal B}(B_c^+ \to h_c(2P)\, \rho^+)/{\cal B}(B_c^+ \to \chi_{c1}(2P) \, \rho^+)$ (first row, middle) and ${\cal B}(B_c^+ \to h_c(2P)\, \rho^+)/{\cal B}(B_c^+ \to \chi_{c2}(2P) \, \rho^+)$ (first row, right panel). The plots in the second row refer to  $K^{*+}$ in the final state. The meaning of the symbols is the same as in Fig.\ref{Fig1}.}\label{Fig2}
\end{center}
\end{figure}

%
%
The conclusion is that the production of the state $\chi_{c1}$ is suppressed, and of the state $h_c$ is enhanced compared to the other charmonia. The effect of changing the   $h_c(2P)$ mass is shown in Figs.~\ref{Fig1} and \ref{Fig2}.


A comment on the mass of  $\chi_{c0}(2P)$ is in order.  We use the mass of  $\chi_{c0}(3860)$ even though another state, $X(3915)$ with $J^{PC}=0^{++}$ has been observed 
and there are  hints on its charmonium nature \cite{Belle:2023yvd}. Our choice respects the hyperfine splitting hierarchy, which would be violated if $X(3915)$ is identified with $\chi_{c0}(2P)$.

\section{Comment on a recent LHCb measurement}\label{LHCb}

The LHCb collaboration has  measured  the ratio of the branching fractions of  nonleptonic  $B_c^+ \to B_s^0 \pi^+$ and $B_c^+ \to J/\psi \pi^+$ decays \cite{LHCb:2022htj}:
\bea
\mathcal{R} = \frac{\mathcal{B}(B_c^+ \to B_s^0 \pi^+)}{\mathcal{B}(B_c^+ \to J/\psi \pi^+)}= 91 \pm 10 \, (stat) \pm 8 \, (syst) \pm 3 \, (\mathcal{B}),
\label{eq:measure}
\eea
where the last uncertainty is due to the knowledge of the branching fractions of the intermediate state decays.
Using naive factorization for the two amplitudes we get

\bea
\Gamma (B_c^+ \to B_s^0 \pi^+) &=& \Lambda \frac{a_1(m_c)^2}{a_1(m_b)^2} \left[f_0(m_\pi ^2)\right]^2 f_{\pi}^2 \lvert V_{cs} \rvert ^2 \lvert V_{ud} \rvert ^2 \left(m_{B_c}^2 -m_{B_s^0}^2 \right)^2 \lambda^{\frac{1}{2}}(m_{B_c}^2,m_{B_s^0} ^2,m_{\pi}^2), \label{eq:GammaBsPi} \nn \\
\Gamma (B_c^+ \to J/\psi \pi^+) &=& \Lambda \left[A_0(m_\pi ^2)\right]^2 f_{\pi}^2 \lvert V_{cb} \rvert ^2 \lvert V_{ud} \rvert ^2 \lambda^{\frac{3}{2}}(m_{B_c}^2,m_{J/\psi} ^2,m_{\pi}^2), 
\label{eq:GammaJPsiPi}
\eea
\noindent where $f_0$ and $A_0$ enter in the parametrization of  $B_c \to B_s$ and $B_c \to J/\psi$ matrix elements, respectively. 
The ratio is:

\bea
\mathcal{R}=\frac{a_1(m_c)^2 \left[f_0( m_\pi^2)\right]^2 \lvert V_{cs} \rvert ^2 \left(m_{B_c}^2 -m_{B_s^0}^2 \right)^2 \lambda^{\frac{1}{2}}(m_{B_c}^2,m_{B_s^0} ^2,m_{\pi}^2)}{a_1(m_b)^2 \left[A_0(m_\pi^2)\right]^2 \lvert V_{cb} \rvert ^2 \lambda^{\frac{3}{2}}(m_{B_c}^2,m_{J/\psi} ^2,m_{\pi}^2)}. \label{eq:RLHCb}
\eea
\noindent
Using the parameters in Table~\ref{tab:input} and  setting $a_1(m_c)=1.089$, the ratio \eqref{eq:RLHCb} depends only on the form factors, which are determined, e.g.,  
by lattice QCD computations, QCD sum rules,  quark models, approaches based on  nonrelativistic QCD (NRQCD).

\noindent The result $f_0(m_{\pi}^2) = 0.625 \pm 0.010$ from lattice QCD \cite{Cooper:2020wnj}, allow us to obtain 
 $\mathcal{B}(B_c^+ \to B_s^0 \pi^+) = 0.0348 \pm 0.0013$. For the $J/\psi$ transition, different determinations of  $A_0(m_\pi^2)$ produce the results
in Table~\ref{tab:A0}.

\begin{table}[!h]
\vspace{0.3cm}
\centering
\begin{tabular}{l cccc}
\hline
& lattice QCD \cite{Harrison:2020gvo} &  relativistic quark model \cite{Colangelo:1999zn} & NRQCD (NLO) \cite{Qiao:2012hp} \\
\hline
$A_0(m_\pi^2)$ & $0.478 \pm 0.031$ & $0.449$ & $0.947$\\
\hline
$\mathcal{R}$ & $46.252^{+6.530}_{-6.464}$ &$48.315$ &$10.871$ \\
\hline
$\mathcal{B}(B_c^+ \to J/\psi \pi^+)$ &$(7.6 \pm 1.0) \times 10^{-4}$ &$6.9 \times 10^{-4}$ &$29.7 \times 10^{-4}$ \\
\hline
\end{tabular}
\caption{\small $B_c \to J/\psi$ form factor $A_0(m_{\pi}^2)$ from different models, and corresponding values of $\mathcal{R}$ and $\mathcal{B}(B_c^+ \to J/\psi \pi^+)$.}\label{tab:A0}
\end{table}
\noindent All methods fail to reproduce  Eq.~\eqref{eq:measure}, the smallest result corresponding to NRQCD. In the improved relativistic quark model \cite{Chang:2014jca} the obtained  $\mathcal{R}$ is also similar to the one in Table~\ref{tab:A0}.
On the other hand, for the  $J/\psi$ modes  the  ratio

\bea
\mathcal{R}_{J/\psi} = \frac{\mathcal{B}(B_c^+ \to J/\psi \pi^+)}{\mathcal{B}(B_c^+ \to J/\psi \mu^+ \nu)}= 0.0469 \pm 0.0028 \pm 0.0046
\label{eq:RJPsi}
\eea
 is measured \cite{LHCb:2014rck}.
Using the form factors in \cite{Harrison:2020gvo} we obtain  $\mathcal{R}_{J/\psi} = 0.0493$. Therefore,  the discrepancy between the values in Table~\ref{tab:A0} and the measurement in \eqref{eq:measure} should be attributed to the $B_s^0$ channel.

The  ratio analogous to  Eq.~\eqref{eq:measure} with a kaon in the final state,   using  lattice QCD form factors, gives

\bea
\mathcal{B}(B_c^+ \to B_s^0 K^+) &=& (2.62 \pm 0.09) \times 10^{-3} , \\
\mathcal{B}(B_c^+ \to J/\psi K^+) &=& 0.05(7)_{-8}^{+8} \times 10^{-3} \label{eq:BRJPsiK} , 
\eea
and
\bea
\mathcal{R}_{K} = \frac{\mathcal{B}(B_c^+ \to  B_s^0 K^+)}{\mathcal{B}(B_c^+ \to J/\psi K^+)} = 45.820^{+6.315}_{-6.248} \,\, . 
\eea
\noi
 Combining  $\mathcal{B}(B_c^+ \to J/\psi K^+)$ and $\mathcal{B}(B_c^+ \to J/\psi \pi^+)$ we have

\bea
\mathcal{R}_{J/\psi}^{NL} = \frac{\mathcal{B}(B_c^+ \to J/\psi K^+)}{\mathcal{B}(B_c^+ \to J/\psi \pi^+)}= 0.0761_{-0.0146}^{+0.0147} \, \, , 
\label{eq:RNLJPsi}
\eea
\noindent in agreement with the measurement  $\mathcal{R}_{J/\psi}^{NL} = 0.079 \pm 0.007 \pm 0.003$ \cite{LHCb:2016vni}.  

The conclusion is that  factorization works quite well for the $J/\psi$ channel.  Deviations emerge in the $B_s^0$ channel,
for which  the argument of colour transparency fails due to the small momentum of pion and kaon in the final state.

\section{Conclusions}\label{end}

Applying naive factorization we have analyzed  the nonleptonic $B_c$ decays to the $P$-wave lowest-lying and    first radial excitations of the charmonium,  and  $\pi^+$, $K^+$, $\rho^+$ and $K^{*+}$.  Using the LO relations among the $B_c \to \chi_{cJ}(h_c)$ form factors  \cite{Colangelo:2022awx}, several ratios of branching fraction are predicted. The $\chi_{c1}$ channel, both for the $1P$ and $2P$ state, is  always suppressed: this should be observed if  $\chi_{c1}(3872)$ is an ordinary  charmonium. On the other hand,  $h_c$ production is enhanced compared to the other states in the multiplet. The results are model-independent, as they are obtained by an expansion of  the form factor expressions.

 An analysis of the recent LHCb measurement in Eq.~\eqref{eq:measure} shows that  the $B_c$  branching fractions involving $J/\psi$ are 
reproduced,   while for $B_c \to B_s \pi^+$ the nonfactorizable contributions are sizeable, as it could be expected on the basis of the color transparency argument.

\vspace*{1.cm}
\noindent {\bf Acknowledgements.}
I  thank Pietro Colangelo, Fulvia De Fazio, Francesco Loparco and Martin Novoa-Brunet for  discussions. This study has been  carried out within the INFN project (Iniziativa Specifica) QFT-HEP.

\bibliography{BcToPWaveH}
\bibliographystyle{JHEP}
\end{document}